# Layer-selective spin-orbit coupling and strong correlation in bilayer graphene

Anna M. Seiler[1], Yaroslav Zhumagulov[2], Klaus Zollner[2], Chiho Yoon[3], David Urbaniak[1], Fabian R. Geisenhof[4], Kenji Watanabe[5], Takashi Taniguchi[6], Jaroslav Fabian[2], Fan Zhang[3], R. Thomas Weitz[1]*

[1] 1st Physical Institute, Faculty of Physics, University of Göttingen, Friedrich-Hund-Platz 1, Göttingen 37077, Germany

[2] Institute for Theoretical Physics, University of Regensburg, 93053 Regensburg, Germany

[3] Department of Physics, University of Texas at Dallas, Richardson, TX, 75080, USA

[4] Physics of Nanosystems, Department of Physics, Ludwig-Maximilians-Universität München, Geschwister-Scholl-Platz 1, Munich 80539, Germany

[5] Research Center for Electronic and Optical Materials, National Institute for Materials Science, 1-1 Namiki, Tsukuba 305-0044, Japan

[6] Research Center for Materials Nanoarchitectonics, National Institute for Materials Science, 1-1 Namiki, Tsukuba 305-0044, Japan

*Corresponding author. Email: thomas.weitz@uni-goettingen.de

**Abstract**

Spin-orbit coupling (SOC) and electron-electron interaction can mutually influence each other and give rise to a plethora of intriguing phenomena in condensed matter systems. In pristine bilayer graphene, which has weak SOC, intrinsic Lifshitz transitions and concomitant van-Hove singularities lead to the emergence of many-body correlated phases. Layer-selective SOC can be proximity induced by adding a layer of tungsten diselenide (WSe$_2$) on its one side. By applying an electric displacement field, the system can be tuned across a spectrum wherein electronic correlation, SOC, or a combination of both dominates. Our investigations reveal an intricate phase diagram of proximity-induced SOC-selective bilayer graphene. Not only does this phase diagram include those correlated phases reminiscent of SOC-free doped bilayer graphene, but it also hosts unique SOC-induced states allowing a compelling measurement of valley *g*-factor and a seemingly impossible correlated insulator at charge neutrality, thereby showcasing the remarkable tunability of the interplay between interaction and SOC in WSe$_2$ enriched bilayer graphene.

**Main**

Various distinct strongly interacting states often compete for the ground state of a clean two-dimensional (2D) electron system under electron-electron interactions. Recently, the seemingly well-explored family of few-layer graphene has gained revived interest [1–10]. For example, in the simplest AB-stacked bilayer graphene (BLG), a plethora of many-body states has been revealed [2,4,5,10]. Here, electric displacement field-controlled inversion symmetry breaking further flattens the bands yet making the low-density saddle points experimentally accessible. Consequently, Stoner ferromagnetic states [2,4,5,10], correlated (semi-) insulating and metallic states [4,5], and superconducting states [2] have been identified close to the field-controlled van-Hove singularities. The detailed nature and mechanisms of these states are under active investigations. For example, it has been shown that proximity-induced spin-orbit coupling (SOC) of Ising (valley-Zeeman) type allows the observation of superconducting states in a much broader parameter space [11,12], leading to the question of whether SOC is responsible for suppressing interacting states that compete with superconductivity [13], or whether superconductivity is enhanced by SOC [12]. To date which effect (e.g., SOC, Coulomb interaction, saddle points) favors which ground state has still been an open question both from the theoretical [13] and experimental [11,12] sides.

While previous investigations have focused on the impact of proximity-induced SOC on the superconductivity of BLG [11,12], impacts of such a SOC on the rest of its correlated phase diagram remains unexplored. Here, we systematically disentangle the intricate interplay between electronic correlations and SOC in spin-orbit proximitized BLG, aiming to elucidate the entire phase diagram. To this end we prioritize the competition between SOC and Coulomb interaction in their impacts on those BLG ground states enriched by the displacement field-controlled saddle points yet outside of the superconducting regime.

A typical device scheme is shown in Figure 1a, where a monolayer of $WSe_2$ is placed below a BLG flake. The BLG/$WSe_2$ heterostructures are contacted by two-terminal graphite flakes and encapsulated in hexagonal boron nitride (hBN). The use of graphite gates ensures the highest device quality. Optical microscope images of the two encapsulated devices investigated in this study are displayed in Fig. 1b and Extended Data Fig. 1.

The dual-gate structure enables independent tuning of the charge carrier density ($n$) and the electric displacement field ($D$). In such a device configuration, Ising SOC (valley-Zeeman type) [14] is expected to only be introduced to the bottom graphene layer [12,15,16], which can be confirmed by our band structure calculations (Fig. 1c) . The proximity-induced SOC results in out-of-plane spin-valley locking of the Ising SOC (valley-Zeeman type), and relatively large (meV) spin-splitting of the low-energy bands of BLG at the K and K' points with no overall layer asymmetry [16]. At $D > 0$ and $n > 0$ (or $D < 0$ and $n < 0$), electronic states near the Fermi energy are polarized to the layer adjacent to the $WSe_2$ [12,15,16], exhibiting the proximity spin-orbit physics. Conversely, the energy bands for $D > 0$ and $n < 0$ (or $D < 0$ and $n > 0$) are only weakly affected by the presence of $WSe_2$ (Fig. 1c). This $D$-dependent spin-valley-locked, spin-split band structure has a direct consequence for the charge transport at small $n$ and $D$ as shown in Fig. 1d, where the occupation of the high-energy spin-split conduction and valence bands, as marked with arrows, are featured by increased $dG/dn$ with conductance $G$ (see Extended Data 3 for a complete phase diagram of both devices investigated for the present study) [15]. The strength of the Ising SOC $\lambda_I$ can be determined directly by mapping the magnetic field dependence of particular Landau-Levels (LL), e.g., the $\nu = 3$ state, to be $\lambda_I = 1$ meV (see Methods and Extended Data Fig. 2), which is consistent with previous results [11,12,15] and DFT calculations [14,16].

**Correlated phases in the vicinity of tunable van-Hove singularities**

We start our analysis in the conduction band at large $D$ fields, where, in the absence of SOC, a cascade of correlated phases was found as low-temperature ground states [5]. Since BLG is only proximitized on one side by $WSe_2$, we can investigate the presence and absence of SOC in the same sample, by selectively addressing the layer-polarized energy levels. We first re-investigate the conduction band without SOC ($D<0$, $n>0$). Consistent with previous works [5,10], three phases can be identified at large $|D|$ close to the band edge: a spin- and valley-polarized (quasi-) insulating (svi) phase, a spin-polarized (quasi-) insulating (si) phase, and a fully spin- and valley-degenerate metallic (m) phase (Fig. 2a-c). The (quasi-) insulating character of the svi and si phases was discussed in Ref. [5]. The different spin and valley degeneracies are identified in out-of-plane and in-plane magnetic field ($B_\perp$ and $B_\parallel$) measurements [5,10] as shown in Fig. 2b and c. Applying a finite $B_\parallel$ (Fig. 2c) that only couples to the spin degree of freedom does not change the phase boundary between the svi and si phases, indicating similar spin polarizations of both phases with a significant in-plane component. By contrast, the phase boundary between the si and m phases shifts towards higher electron densities, further indicating spin polarizations in the svi and si phases but not in the m phase. Applying $B_\perp$ that couples to the valley and out-of-plane spin degrees of freedom via the orbital and spin Zeeman effects affects both phase boundaries (Fig. 2b), indicating a change in valley polarization between the svi and si phases.

A drastically different phase diagram appears in the same parameter space ($|D|$ and $n$) but for the WSe$_2$ proximitized conduction band ($D>0$, $n>0$, Fig. 2d-f). We do not observe signatures of Stoner or other interacting phases, but only the SOC-induced splitting of the originally four-fold degenerate band into two sets of two-fold degenerate spin-valley-locked bands. While the transition between the two sets of SOC-split bands shown in Fig. 1d extends to large $D$ (Fig. 2d), in the investigated range up to 0.8 V/nm, no additional conductance features indicating additional spin or valley polarization appear at $D > 0$ and $B = 0$, unlike on the non-WSe$_2$-proximitized side. It appears that in the presence of SOC, the formation of spin and/or valley polarized phases (i.e., si and svi) near the conduction band edge of $D$ field-gapped BLG is suppressed. This picture is also consistent with the magnetic field data. Applying $B_\perp$ leads to a valley splitting of the low-energy spin-valley-locked band due to the valley-Zeeman effect, as indicated by a peak in the d$G$/d$n$ marked with a dashed line in Fig 2e. From the $B_\perp$-field induced crossing (marked as $B_C$ in Extended Data Fig. 4) between the upper band of the lower set and the lower band of the upper set, which have the same spin but opposite valley indices, we can determine the *D*-dependent valley g-factor $g_V$, which quantifies the strength of the valley magnetic moment [17]. Using $\lambda_I = \frac{1}{2} \mu_B g_V B_C$ with Bohr magneton $\mu_B$ and $B_C$ = 1.9 T, we find $g_V$ to be 22 at *D* = 0.7 Vnm$^{-1}$, which aligns well with previous results obtained in BLG quantum dots [18–20]. In the $B_{||}$ map up to 2T we do not observe any change of the phase boundaries, since in this regime the spin Zeeman energy $E_z \ll \lambda_I$ [16].

All our observations point to the absence of correlated phases when the electrons are polarized to the layer adjacent to WSe$_2$. Theoretically, the Ising SOC of out-of-plane spin quantization, together with the WSe$_2$-enhanced screening effect, is expected to disfavor those correlated phases with in-plane spin orientations as previously identified in the absence of SOC [4,5]. We note that WSe$_2$ does have a larger dielectric constant than hBN [21]. Given that monolayer WSe$_2$ is used in our devices, the short-range part of Coulomb interaction is more significantly screened than the long-range Coulomb tail. Nevertheless, – even though the effect will likely be small – the WSe$_2$-enhanced screening may disfavor Stoner and more elaborate correlated phases [22–24].

The picture becomes more complex on the hole-doped side of BLG, where trigonal warping is more pronounced [25] and Lifshitz transitions with concomitant van-Hove singularities give rise to a cascade of correlated insulating and metallic phases of both Stoner and non-Stoner types [4]. These phases were initially observed in pristine BLG and can also be identified similarly in the BLG/WSe$_2$ devices for *D* > 0 and *n* < 0, where charge transport occurs in the layer non-adjacent to WSe$_2$ (see Fig. 3a,b). However, the proximity-induced Ising SOC is present for *D* < 0 and *n* < 0 when charge transport occurs in the layer adjacent to WSe$_2$, and evidently it disfavors these correlated phases (see Fig. 3c-f for the data from Device II and Extended Data Fig. 5 for data from Device I). Notably, our findings reveal multiple distinctive features in the density derivative of the conductance, as indicated by arrows in Figure 3d, which is in sharp contrast to our observations at low *D* fields (Fig. 1d), at electron doping (Fig. 2d-f), and in previous literature [13].

The first feature at lowest $n$ near the band edge separates low and high conductance regimes (red arrows in Fig. 3c,d). In the low-conductance regime, the conductance sharply increases with increasing applied current $I_{Bias}$ (Extended Data Fig. 6a,b), indicating insulating behavior, possibly arising from disorder [26,27], Wigner crystallization [28], or more exotic phenomena, e.g., induced by exchange interaction between the trigonal warping induced mini valleys [29–31]. A detailed analysis of this regime, however, extends beyond the scope of this present study. This regime is markedly different from that when WSe$_2$ is absent, where a metallic phase had been observed at the valence band edge [4].

Another marked difference is in the higher density regime, where a cascade of correlated metallic and insulating phases were found [4] in the absence of SOC, whereas in the presence of SOC no clear signs of physics beyond SOC can be identified. These features can be associated with the SOC-induced band splitting and the occupation of the second energy band (green arrow in Fig. 3c-f), exhibiting $B_{||}$- and $B_{\perp}$-dependences similar to those observed at the electron doped case with induced SOC (Extended Data Fig. 7, Fig. 3e,f). This identification is reinforced by the appearance of additional Landau levels that curve in the $n$-$B_{\perp}$ space (dashed lines in Fig. 3e). Since trigonal warping is more pronounced in the valence band [25], changes in the conduction can further be observed when the Fermi energy is tuned to the vicinity of Lifshitz transitions (grey triangles in Fig. 3f). Such features in the conductance have been captured in a continuum model of BLG with the addition of an Ising SOC term [15] and, however, escaped previous experimental observations. With the exception in this 'triangle' phase of the first band, the prevalence of mainly SOC-driven states is also supported by the absence of non-linear d$I$/d$V$ feature and the lack of strong temperature dependence (Extended Data Fig. 6a-c). Finally, we note the absence of any indication of superconductivity near these potential phase boundaries [11,12]. This may be attributed to constraints imposed by the application of higher $D$-fields due to elevated gate currents and the use of thicker hBN dielectrics in the range of 37 nm to 62 nm compared to two previous studies [11,12].

**Transport in WSe$_2$ proximitized BLG at overall charge neutrality**

While at large *D* fields and finite doping the correlated phases are strongly suppressed when charge carriers are layer-polarized to the WSe$_2$-BLG interface, the situation is different at charge neutrality when there is no layer polarization (*D* = *n* = 0). The ground state of undoped BLG has been intensively addressed by multiple works [7,32–36]; the strength of Coulomb interaction plays a critical role. In the limit of strong interaction, as experimentally realized in suspended BLG, an insulating layer antiferromagnetic (LAF) state has been identified as the ground state [7,32–36]. In the LAF state, two spin species spontaneously polarize to opposite layers [17]. On the other hand, in hBN encapsulated BLG, interaction is weaker due to the larger dielectric constant $\epsilon_r$ and closer distance of screening metal gates (about 150 nm in the suspended samples and typically below 60 nm in hBN encapsulated samples), and the LAF state has not been found. Thus, one may wonder whether the ground state is still correlated or non-interacting given the induced SOC but the weakened Coulomb interaction.

From our multi-scale theoretical investigation of the single-particle bands (Fig. 4f,g), consistent with previous reports, a SOC-split but overall gapless band structure is present. An interesting question now is whether the induced Ising SOC and the more screened interaction lead to a correlated insulating state like in the freestanding case or disfavor correlated states as in the case of large *D* fields discussed above. Figure 4 shows our experimental observations, and a region of suppressed conductance is evident at charge neutrality. This region is unstable against the application of doping or a *D* field of either sign. These suggest most likely the topologically trivial LAF state as in the freestanding case [7,32–36], and with increasing $B_\perp$ the LAF state evolves into a canted antiferromagnetic (CAF) state (Fig. 4c) [37–39]. Furthermore, this state is not strongly affected by the application of a $B_{||}$ field (Fig. 4d), consistent with two previous studies [36,39]. In addition, we observe insulating behavior, i.e., decreasing resistance with increasing temperature or increasing current (Fig. 4e and Extended Data Fig. 8). The insulating state is stable below 5 K, and we extract an energy gap of 0.4 meV using Arrhenius fits (see inset of Fig. 4e and Methods). Our observations are not only at odds with the single-particle band structure of our WSe$_2$/BLG heterostructure but also at odds with a previous study of WSe$_2$/BLG structures [15]. (We note that a region of decreased conductance was also observed in a WSe$_2$/BLG/ WSe$_2$ structure, where symmetrically induced Ising SOC results in a single-particle Kane-Mele SOC gap that is suppressed by the application of a $B_{||}$ field [15]). The respective stability of the LAF state in the D, $B_\perp$, $B_{||}$ space is consistent with the observations made in freestanding graphene (see Extended Data Fig. 9 for a direct comparison of activation gaps). These experimental observations also align with our self-consistent Hartree-Fock calculations addressing both short- and long-range interactions (see methods for details): opening a spontaneous gap in a BLG/WSe$_2$ heterostructure requires a smaller Coulomb interaction (Fig. 4f,g), and the larger the induced SOC the stronger the spontaneous indirect gap opening (Fig. 4h).

**Conclusions**

In summary, asymmetrically introducing WSe$_2$ to hBN-encapsulated BLG devices provides a unique platform for the distinct exploration of correlated phases and SOC-induced states, respectively, in the presence of large *D* fields, depending on the signs of *n* and *D*. Whereas a cascade of correlated phases emerges, like those devices without WSe$_2$, when the charge carriers are polarized on the WSe$_2$-remote layer of BLG, clear signatures of SOC dominant band splitting have been identified allowing the further measurement of a strong valley *g*-factor, as if the interaction is nearly absent, when the charge carriers are polarized on the WSe$_2$-proximate layer of BLG. At zero *D* field for which the BLG is not layer-polarized, surprisingly, the interaction strength in BLG appears to be enhanced by the induced SOC, giving rise to a correlated insulating state with both theoretically and experimentally anticipated features under $B_\perp$, $B_{||}$, and *D* fields. Our results have established a rich phase diagram of BLG, paving the way for exploring the interplay between geometry, interaction, and SOC in strongly correlated electrons.

**Note from the authors**

While analyzing the data we became aware of similar results presented in a manuscript by Masseroni et al. It is remarkable that very similar data was obtained by two different groups, using a different TMD on bilayer graphene (WSe$_2$ in Göttingen and MoS$_2$ in Zurich).

**Methods**

**Device fabrication**

BLG flakes, graphite flakes and hBN flakes were exfoliated on Si/SiO$_2$ substrates and characterized using optical microscopy, Raman spectroscopy and atomic force microscopy. In a parallel procedure, WSe$_2$ monolayers were directly exfoliated onto a silicone gel-film (DGL Film, Gel-Pak) and identified through optical microscopy and Raman spectroscopy.

Devices I and II were assembled in a four-step process. First, the bottom hBN flake and bottom graphite flake were stamped on a clean Si/SiO$_2$ substrate, employing a dry transfer method that is described in detail in Reference [40]. Subsequently, a WSe$_2$ monolayer was transferred from a silicone gel-film onto the previously prepared stack. Following this, a top hBN, two graphite contacts and a BLG flake were stamped on top, without any specific twist angle being introduced between the WSe$_2$ and BLG flakes. The graphite top gate was then transferred in a final stamping step. After each stamping iteration, the samples were annealed at 300 °C for 12 h.

The thicknesses of the hBN flakes, serving as dielectrics, were determined to be 58 nm (top dielectric of Device I), 62 nm (bottom dielectric of Device I), 58 nm (top dielectric of Device II), and 37 nm (bottom dielectric of Device II) by atomic force microscopy. Metal contacts made from chromium and gold, connecting the graphite contacts and gates with larger pads, were structured using electron-beam lithography and evaporated onto the sample. Optical images, as well as a schematic representation of our devices, are presented in Fig. 1 and Extended Data Fig. 1.

Device III, whose data is showcased in Extended Data Fig. 9, is a freestanding bilayer graphene device that had been characterized in Ref. [7] in greater detail.

**Electrical measurements**

The electrical measurements were carried in a dilution refrigerator featuring a superconducting magnet. Unless specified otherwise, the sample temperature was maintained between 10 and 20 mK.

The two-terminal conductance that includes a contact resistance was obtained using an AC current ranging from 1 to 20 nA at a frequency of 78 Hz, employing Stanford Research Systems SR865A lock-in amplifiers. DC bias currents $I_{Bias}$ were added through a Keithley 2450 SourceMeter where mentioned in the main text. Home-made low-pass filters were employed to minimize high-frequency noise. Gate voltages were applied through Keithley 2450 SourceMeters.

The charge carrier density *n* and electric displacement field *D* were applied by independent tuning of the top and bottom gate voltages ($V_t$ and $V_b$). They are defined as

$$n = \frac{1}{e}(C_t V_t + C_b V_b)$$

and

$$D = \frac{1}{2\varepsilon_0}(C_t V_t - C_b V_b),$$

where $\varepsilon_0$ is the vacuum permittivity, and $C_t$ and $C_b$ are the top-gate and bottom-gate capacitances. $C_t$ and $C_b$ were extracted at low *D*-fields by aligning the integer quantum Hall plateaus at finite magnetic fields with their corresponding slopes in a Fan diagram.

**Determining the strength of SOC**

We determine the strength of the Ising SOC, denoted as $\lambda_I$, by examining the zeroth Landau level octet at positive and negative *D*-fields, following the methodology outlined in [15]. The two sets of Landau levels appearing at positive and negative *n* have opposite layer polarization. The layer-selective SOC results in a $B_\perp$- dependent energy difference $\Delta E = E_Z + \lambda_I/2$ between them, where $E_Z = g\mu_B B_\perp$ is the Zeeman energy with g-Factor g and Bor magneton $\mu_B$. $\Delta E$ manifests as an offset in the Landau level crossings at filling factors $\nu = \pm 3$. This offset vanishes at a critical magnetic field $B^*$ where $2E_Z = 2g\mu_B B^* = \lambda_I$. $B^*$ and $\lambda_I$ can thus be determined by identifying $B^*$. Tracing down the $D$ values at which Landau level crossings occur for different $B_\perp$ (see arrows in Extended Data Fig. 2a-c) enables to extrapolate $B^*$ (Extended Data Fig. 2d). For Device II, $B^*$ was determined to be 4.6 T, resulting in a SOC strength of $\lambda_I = 1$ meV.

**Arrhenius fits**

The activation gaps $\Delta$ depicted in the inset of Fig. 4e and in Extended Data Fig. 9 were determined using the Arrhenius activation model $R = R_0 \, e^{\Delta/2k_B T}$, where $R_0$ is a constant, $k_B$ is Boltzmann's constant and $T$ is temperature. $\Delta$ was extracted for constant $D$ by fitting the resistance $R$ within a temperature range spanning from 1.65 K to 10 K (see Fig. 4e for the corresponding resistances).

**Multi-scale modeling of BLG/WSe₂ heterostructure**

**Structural setup**

The BLG/WSe₂ heterostructure is set up with the atomic simulation environment (ASE) [41] and the CellMatch code [42], implementing the coincidence lattice method [43,44]. We employ a 4 x 4 supercell BLG, stacked on top of a 3 x 3 supercell of WSe₂. The twist angle between the layers is 0°. The graphene lattice constant in the heterostructure is 2.46 Å, and the WSe₂ lattice constant is 3.28 Å. Therefore, the individual monolayers are barely strained in our heterostructure, allowing us to reliably extract band offsets as well as proximity effects on the low energy states of BLG. In order to simulate quasi-2D systems, we add a vacuum of about 16 Å to avoid interactions between periodic images in our slab geometry. The resulting heterostructure is shown in Extended Data Fig. 10.

**First-principles calculations**

The electronic structure calculations and structural relaxations of the BLG/WSe$_2$ heterostructure are performed by density functional theory (DFT) [45] with Quantum ESPRESSO [46]. Self-consistent calculations are carried out with a *k*-point sampling of 12 x 12 x 1. The energy cutoff for charge density is 560 Ry, and the kinetic energy cutoff for wavefunctions is 70 Ry for the fully relativistic pseudopotentials with the projector augmented wave method [47] with the Perdew-Burke-Ernzerhof exchange correlation functional [48]. For the relaxation of the heterostructures, we add DFT-D2 vdW corrections [49–51] and use quasi-Newton algorithm based on trust radius procedure. Dipole corrections [52] are also included to get correct band offsets and internal electric fields. To get proper interlayer distances and to capture possible moiré reconstructions, we allow all atoms to move freely within the heterostructure geometry during relaxation. Relaxation is performed until every component of each force is reduced to below $5 \times 10^{-4}$ Ry/$a_0$, where $a_0$ is the Bohr radius. The relaxed interlayer distance between the two graphene layers is 3.24 Å, and that between the adjacent graphene and WSe$_2$ layers is 3.36 Å.

**Band structure and density of states**

In Fig. 1c, Fig. 4b and Extended Data Fig. 11 we show the calculated band structure of the BLG/WSe$_2$ heterostructure. The low energy bands of proximitized BLG are localized within the WSe$_2$ band gap and at the heterostructure Fermi level. The low energy bands of BLG exhibit an energy gap of about 10 meV, which is due to the intrinsic dipole of about 0.68 debye of the heterostructure. Additionally, the low energy valence bands are split, due to short-range proximity-induced SOC. The results are similar to previous literature, and the physics has been elucidated in Refs. [16,53,54].

The model Hamiltonian that we employ is adapted from Ref. [54]. We find the following parameters informed by our current DFT calculations: $\gamma_0$ = 2.555 eV, $\gamma_1$ = 0.412 eV, $\gamma_3$ = -0.311 eV, $\gamma_4$ = -0.178 eV, $V$ = -6.965 meV, $\Delta$ = 13.371 meV, $\lambda_{R1}$ = -0.516 meV, $\lambda_{R2}$ = 0 meV, $\lambda_I^{A1}$ = 1.121 meV, $\lambda_I^{B1}$ = -1.192 meV, $\lambda_I^{A2}$ = 0 meV, $\lambda_I^{B2}$ = 0 meV, $E_D$ = 2.220 meV. From the dipole of the heterostructure, we know that an external electric field of about -0.25 V/nm is needed to compensate the built-in electric field. As such the gap fully closes at the compensated $V$ = 0.

Employing the DFT informed model Hamiltonian, we can simulate the response of the low energy bands upon an external electric field. The low energy bands and the corresponding density of states (DOS) is shown in Extended Data Fig. 12, for a series of interlayer electric potentials $V$. The Fermi level corresponds to the dashed lines, i.e., the zero energy. Charge neutrality is at the energy at which electron and hole DOS contributions compensate and is calculated from the band resolved DOS. The evolution of DOS as a function of $V$ at the Fermi level and at charge neutrality is shown in Extended Data Fig. 13. We find that in both cases the DOS decreases with increasing $V$, inconsistent with the experimental data (Figure 4a).

**Self-consistent Hartree-Fock calculations**

We address the roles played by both long-range and short-range Coulomb interactions separately and show both theories are consistent with our experimental observations. The Hartree-Fock theory of using an averaged long-range Coulomb interaction has been well documented in Ref. [34,39] and successfully explained the spontaneous gap opening in BLG and its dependences on temperature, density, and in-plane magnetic field. As detailed in a recent report [55], this theory has incorporated the proximity-induced SOC and average screening effects to elucidate the observation of quantum anomalous Hall effect with a large Chern number $C$ = 5 in ABC pentalayer graphene at charge neutrality [56]. Applying the same theory to the hBN encapsulated BLG/WSe$_2$ heterostructure, we find that the larger the induced SOC the stronger the spontaneous indirect gap opening (Fig. 4h).

We study the short-range Coulomb interaction effect at zero level of doping using the minimal density-density interactions outlined as follows:

$$H_{int} = U_0 \sum_{l \in (t,b)} N_l(N_l - 1)/2 + U_1 N_t N_b,$$

where $N_l$ stands for a layer-resolved number operator with layer index $l$. $U_0$ denotes the intralayer contact Coulomb interaction amplitude, while $U_1$ denotes the interlayer contact Coulomb interaction amplitude. We utilized a self-consistent Hartree-Fock theory to establish the correlated band structure. The effective Schrödinger equation is provided as follows:

$$\sum_b [h_{\bar{a}b}(k) + \hat{\Sigma}_{\bar{a}b}] u_{nk}^b = (\varepsilon_{nk} + \mu) u_{nk}^a,$$

where $h_{\bar{a}b}(k)$ represents the bare kinetic Hamiltonian in bilinear form, defined in spin, valley, and layer degrees of freedom and obtained by numerically downfolding [57] the four-band DFT-informed Hamiltonian. As a result, the bare kinetic Hamiltonian is defined in B1/A2 low energy sublattices, which can be referred to as layer index. $\Sigma_{\bar{a}b}$ denotes the Hartree-Fock self-energy, $u_{nk}$ refers to the single-particle wavefunction, and $\varepsilon_{nk}$ is the single-particle energy. Indexes $\bar{a}b$ correspond to spin, valley, and layer degrees of freedom. The Hartree-Fock self-energy is calculated as follows:

$$\Sigma_{\bar{b}a} = -\sum_{cd} \Gamma_{a\bar{b}c\bar{d}} G_{c\bar{d}},$$

where $\Gamma_{a\bar{b}c\bar{d}}$ is the interaction vertex function and $\hat{G}_{c\bar{d}}$ is the density matrix. The interaction vertex function follows from the interaction Hamiltonian

$$H_{int} = \frac{1}{4} \sum_{abcd} \Gamma_{c\bar{a}d\bar{b}} c_{\bar{a}}^\dagger c_{\bar{b}}^\dagger c_c c_d,$$

and the density matrix is subsequently determined as:

$$G_{c\bar{d}} = A_{uc}\int_{|k|<\Lambda}\frac{d^2k}{(2\pi)^2}\left[f_{nk} - \frac{1}{2}\right]u^c_{nk}(u^d_{nk})^*,$$

where $\Lambda$ is the momentum cutoff. In our calculations, we use a momentum cut-off $\Lambda = 0.06$ Å$^{-1}$. We constrain the chemical potential by fixing electron doping to

$$n_e = \int_{|k|<\Lambda}\frac{d^2k}{(2\pi)^2}\sum_n\left[f_{nk} - \frac{1}{2}\right],$$

Where the offset -1/2 in the density matrix and chemical potential equations is intended to fix the chemical potential equal to zero at the charge neutrality point. On top of the Hartree-Fock calculations, we can obtain the temperature-smeared DOS at the Fermi level as follows:

$$\rho_F = \frac{1}{k_BT}\int_{|k|<\Lambda}\frac{d^2k}{(2\pi)^2}\sum_n\left[4\cosh\left(\frac{\varepsilon_{nk} - \mu}{2k_BT}\right)\right]^{-2}.$$

Using the Hartree-Fock theory, we calculate how the DOS at the Fermi level at zero doping depends on temperature. By setting $U_0$ = 36 eV and $U_1$ = 27 eV, we achieve results (Fig. 4f) consistent with the experimental resistance/temperature/gate phase diagram for BLG/WSe$_2$ heterostructure. Also, we construct the spin-resolved band structure (Extended Data Fig. 14). With the same Coulomb interaction parameters, we do not observe any changes in the band structure for the bare BLG system. Unlike in a bare BLG system, a correlated phase with a spontaneous gap is observed in a proximitized BLG/WSe$_2$ system at zero doping level. For the bare BLG system, we achieve results consistent with the experimental resistance/temperature/gate phase diagram by setting $U_0$ = 68 eV and $U_1$ = 50 eV (to model the freestanding case). We observe a layered antiferromagnetic correlated phase with a spontaneous gap, which can be described using the symmetry-breaking mean-field term:

$$\Phi_{LAF} = \sum_{|\boldsymbol{k}|<k_c}\sum_i c^\dagger_{\bar{a}}(\boldsymbol{k})[M_\Phi]_{\bar{a}b}c_b(\boldsymbol{k}), M_\Phi = l_z \otimes s_z \otimes \tau_0,$$

where $l$, $s$ and $\tau$ are Pauli matrices acting on layer, spin, and valley degrees of freedom, respectively. Opening a spontaneous gap in the proximitized BLG/WSe$_2$ heterostructure requires a smaller amplitude of the Coulomb interaction compared to the same system but without WSe$_2$. Therefore, it is possible to observe a spontaneous gap in the hBN encapsulated BLG/WSe$_2$ heterostructure (in addition to the free-standing bare BLG case) but not in the hBN encapsulated bare BLG case. For comparison, we plot in Fig. 4g the dependence of the spontaneous gap on the amplitude of the Coulomb interaction U = U$_0$ = 1.3 U$_1$.

**Acknowledgements**

We thank Michele Masseroni, Thomas Ihn, and Klaus Ensslin for fruitful discussions. R.T.W. and A.M.S. acknowledge funding from the Deutsche Forschungsgemeinschaft (DFG, German Research Foundation) under the SFB 1073 project B10. R.T.W. acknowledges partial funding from the SPP2244 from the Deutsche Forschungsgemeinschaft (DFG, German Research Foundataion). K.W. and T.T. acknowledge support from the JSPS KAKENHI (Grant Numbers 20H00354, 21H05233 and 23H02052) and World Premier International Research Center Initiative (WPI), MEXT, Japan. K.Z., Y.Z. and J.F. were supported by the Deutsche Forschungsgemeinschaft (DFG, German Research Foundation) SFB 1277 (Project No. 314695032), SPP 2244 (Project No. 443416183), the European Union Horizon 2020 Research and Innovation Program under contract number 881603 (Graphene Flagship) and FLAGERA project 2DSOTECH. The theoretical work done at UT Dallas was supported by NSF under Grants no. DMR-1945351, no. DMR-2105139, and no. DMR-2324033. We acknowledge the Texas Advanced Computing Center (TACC) for providing resources that have contributed to the research results reported in this work.

**Author contributions**

A.M.S. fabricated Device I and II with help of D.U. and conducted the measurements and data analysis. F.R.G fabricated and measured Device III. K.W. and T.T. grew the hexagonal boron nitride crystals. Y.Z, K.Z, C.Y., J.F. and F.Z. contributed to the theoretical part. All authors discussed and interpreted the data. R.T.W. supervised the experiments and the analysis. The manuscript was prepared by A.M.S., Y.Z., K.Z., J.F., F.Z. and R.T.W. with input from all authors.

**Corresponding authors**

R. Thomas Weitz ([thomas.weitz@uni-goettingen.de](thomas.weitz@uni-goettingen.de))

**Competing interests**

Authors declare no competing interests.

**Data availability**

All data supporting the messages of the manuscript is displayed in the manuscript. The raw data is available from the authors upon request.

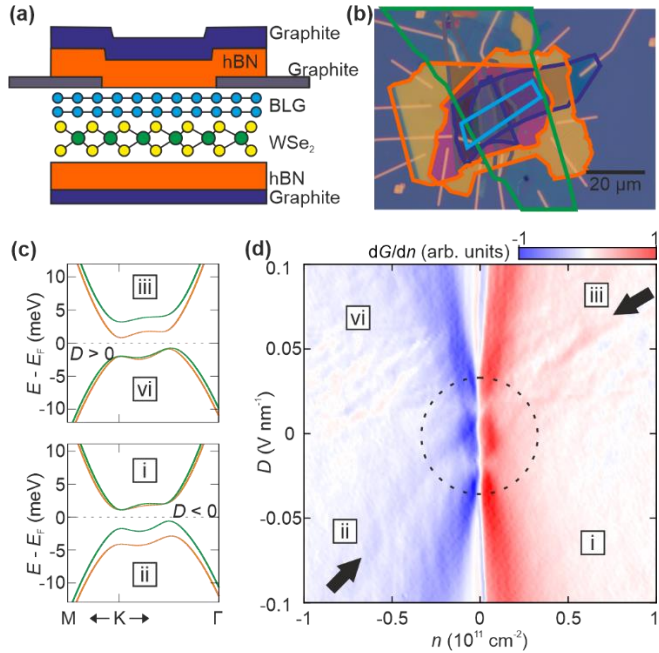

**Fig. 1. Phase Diagram of bilayer graphene exhibiting layer-selective proximity-induced SOC. (a)** Device schematic of an encapsulated bilayer graphene flake located on top of a WSe$_2$ flake. **(b)** Optical microscope image of Device I. Graphite flakes serving as gates (contacts) are outlined in purple (grey), hBN flakes are outlined in orange, the WSe$_2$ flake is outlined in green, and the BLG flake is outlined in blue (the same color code as in **(a)**). **(c)** Low-energy band structures of bilayer graphene near the K-point of the Brillouin zone for electric potentials $V$ = -2 meV (top) and $V$ = 2 meV (bottom), calculated from a model that includes an Ising SOC of $\lambda_I$ = 1.07 meV on the bottom layer. **(d)** Derivative of the two-terminal conductance $G$ (d$G$/d$n$) as a function of charge carrier density $n$ and electric displacement field $D$ measured in Device I with $B_\perp = B_{||}$ = 0 T. Changes in the conductance associated with proximity-induced spin-orbit coupling are marked by arrows or encircled. The labels i-iv mark the four quadrants in the $n$-$D$ plane and correspond to the energy bands shown in **(c)**.

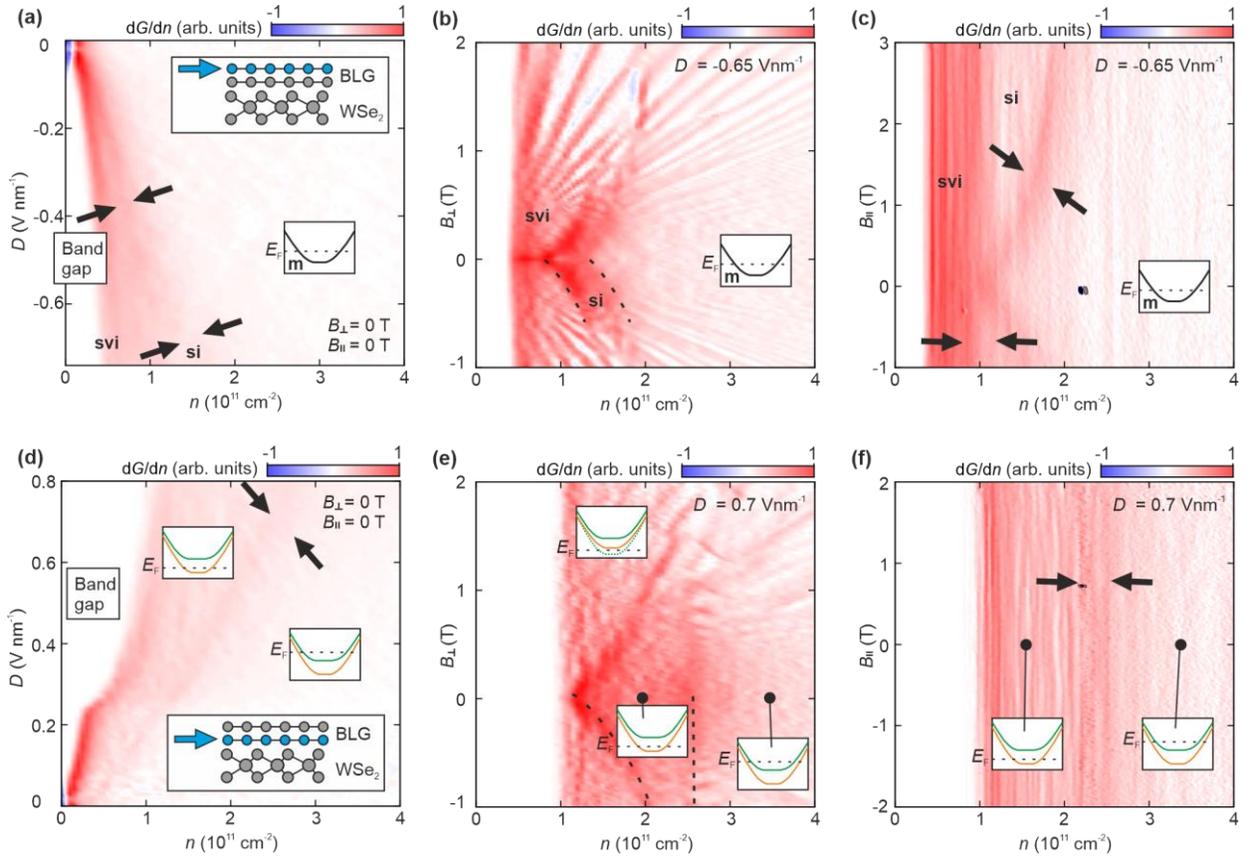

**Fig. 2. Phase diagrams of electron-doped bilayer graphene without and with layer-selective proximity-induced SOC. (a)** d$G$/d$n$ as a function of $n$ and negative $D$ measured in Device I with $B_\perp = B_{||}$ = 0 T. Electron transport is taking place in the top layer where the effects of SOC are weak. The **svi**, **si** and **m** phases are labeled, and phase boundaries are highlighted by arrows. **(b,c)** d$G$/d$n$ as a function of $n$ and $B_\perp$ **(b)** and $B_{||}$ **(c)** for $D$ = -0.65 V nm$^{-1}$. Phase boundaries are highlighted by dashed lines or arrows. **(d)** d$G$/d$n$ as a function of $n$ and positive $D$ measured in Device I with $B_\perp = B_{||}$ = 0 T. Electron transport is taking place in the bottom layer where the effects of SOC are strong. **(e,f)** d$G$/d$n$ as a function of $n$ and $B_\perp$ **(e)** and $B_{||}$ **(f)** for $D$ = -0.65 V nm$^{-1}$. Phase boundaries are highlighted by dashed lines or arrows. The insets schematically show the low-energy band structures in the respective regions of the phase diagrams.

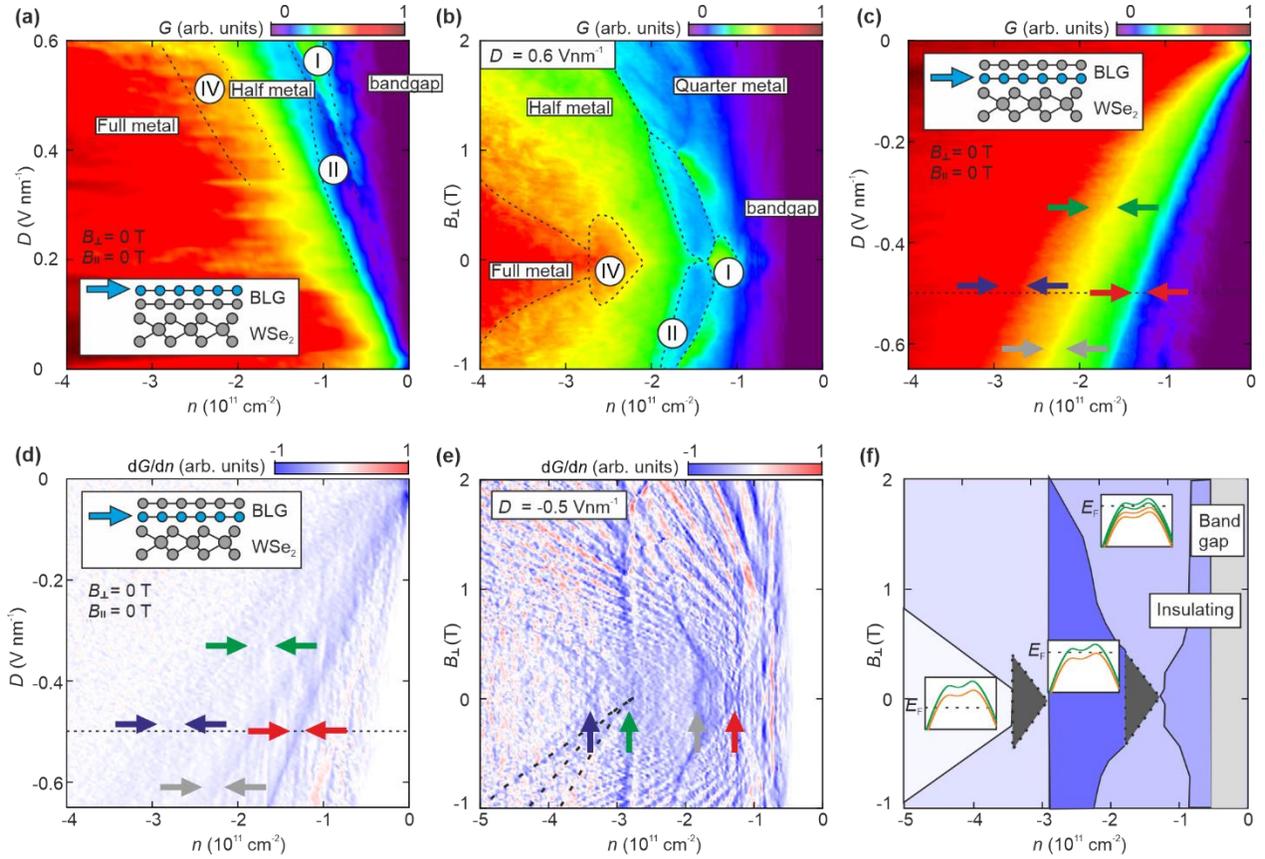

**Fig. 3. Phase diagrams of hole-doped bilayer graphene without and with layer-selective proximity-induced SOC. (a)** Conductance $G$ as a function of $n$ and positive $D$ measured in Device II with $B_\perp = B_{||}$ = 0 T. Electron transport is taking place in the top layer where the effects of SOC are weak. Phase boundaries are highlighted by dashed lines. The interaction-induced phases are labeled according to Ref. [4]. **(b)** $G$ as a function of $n$ and $B_\perp$ at $D$ = 0.6 Vnm$^{-1}$. Correlated phases are labeled using the nomenclature of Ref. [4] where half and quarter metal phases are Stoner ferromagnetic states, phases I and IV correspond to correlated metals of non-Stoner type and phase II corresponds to a correlated insulating phase consistent with a Wigner-Hall crystal. **(c,d)** $G$ **(c)** and d$G$/d$n$ **(d)** as a function of $n$ and negative $D$ measured in Device II with $B_\perp = B_{||}$ = 0 T. Electron transport is taking place in the bottom layer where the effects of SOC are strong. Phase boundaries are highlighted by arrows. **(e)** d$G$/d$n$ as a function of $n$ and $B_\perp$ for $D$ = -0.5 V nm$^{-1}$. Phase boundaries are highlighted by arrows. Quantum Hall states arising from the second energy band are traced by dashed lines for negative $B_\perp$. **(f)** Schematic

of the different phases induced by SOC. The insets show the low-energy band structures in the respective regions of the phase diagrams.

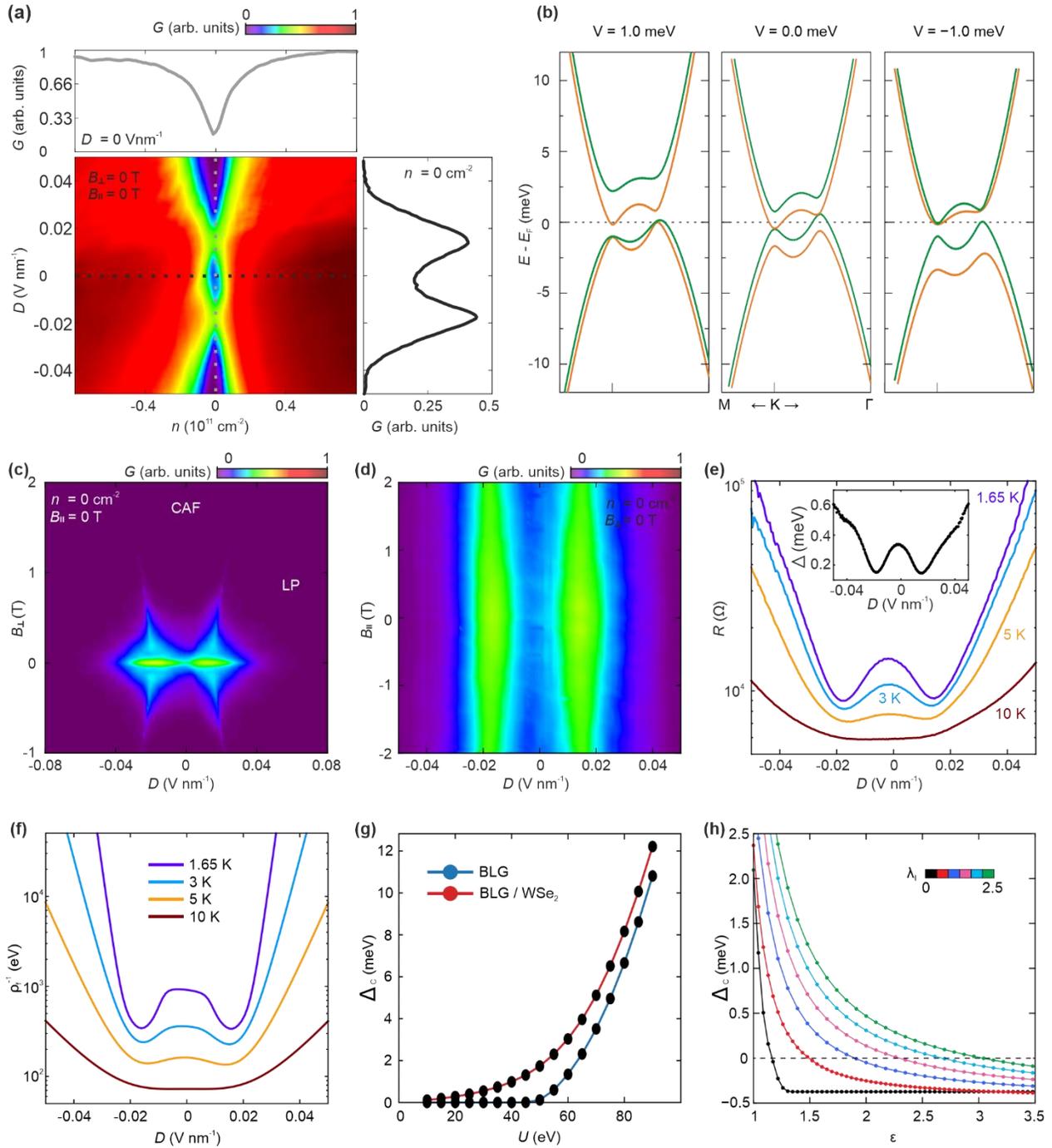

**Fig. 4. Correlated insulating state at $D = n = B_\perp = 0$ and multi-scale modeling. (a)** Two-terminal conductance $G$ in arbitrary units (arb. units) as a function of $D$ and $n$ at zero magnetic field $B$ measured in Device I. Line traces are shown for $D = 0$ and $n = 0$, respectively. **(b)** Non-interacting band structures of bilayer graphene on top of WSe$_2$ at different interlayer electric potentials. No band gap is open at $V = 0$. **(c)** $G$ as a function of $D$ and an out-of-plane magnetic field $B_\perp$ at $n = 0$. The layer antiferromagnetic (LAF), canted antiferromagnetic (CAF) and layer polarized (LP) states are labeled. **(d)** $G$ as a function of $D$ and an in-plane magnetic field $B_\parallel$ at $n = 0$ and $B_\perp = 0$. **(e)** Two-terminal resistance $R$ as a function of $D$ for different temperatures. The inset shows the size of the activation gap Δ, as determined via Arrhenius fits, as a function of $D$. **(f)** Inverse density of states at Fermi level as a function of $D$ for different temperatures calculated from a Hartree-Fock theory for BLG/WSe$_2$ heterostructure at zero level of doping. **(g)** Trend of the dependence of the correlated gap (direct gap) on the amplitude of the onsite Coulomb interaction $U = U_0 = 1.3\, U_1$ for BLG and proximitized BLG/WSe$_2$ systems. **(h)** Trend of the dependence of the long-range Coulomb interaction induced spontaneous gap (indirect gap) on the SOC strength $\lambda_I$ and the dielectric constant $\varepsilon$.

**Extended Data:**

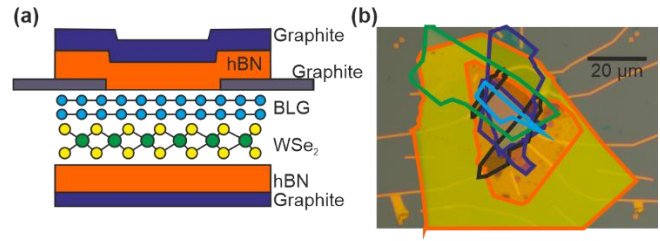

**Extended Data Fig. 1. Device II. (a)** Device schematic of an encapsulated bilayer graphene flake located on top of a WSe$_2$ flake. Graphite contacts are used to contact the bilayer graphene. **(b)** Optical microscope image of Device II. The flakes are outlined using the color code of the device schematic shown in **(a)**.

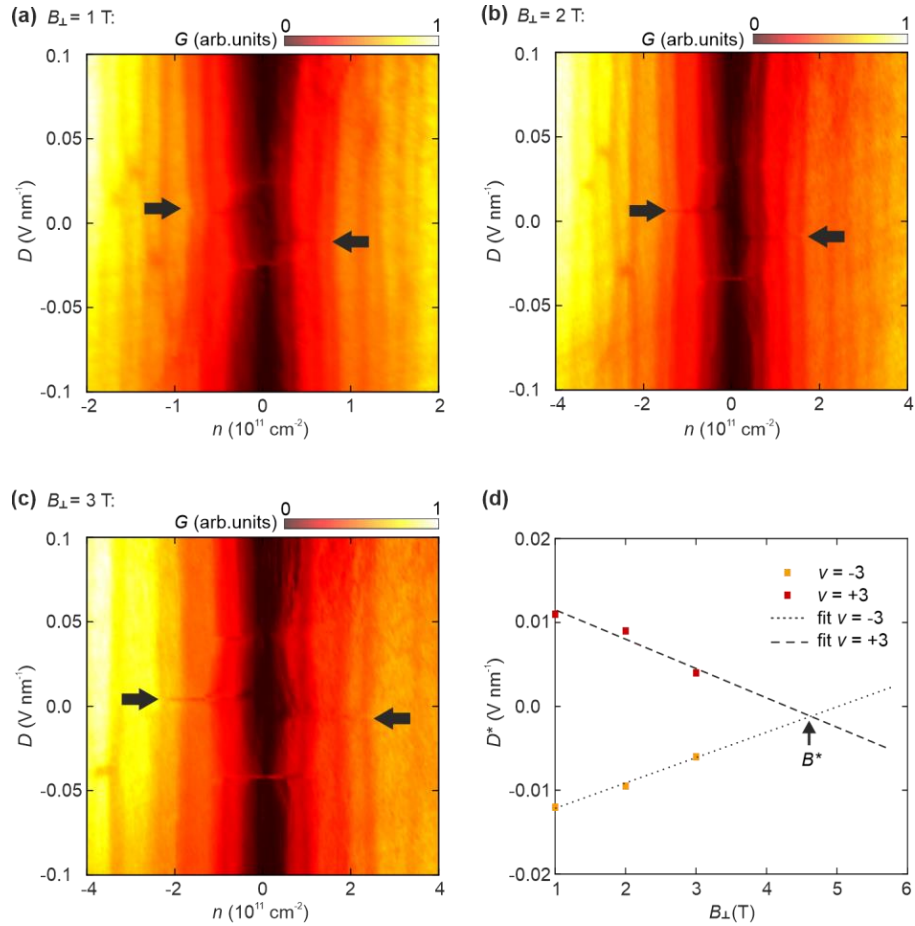

**Extended Data Fig. 2. Determining the Ising SOC in Device II. (a-c)** Two-terminal conductance *G* as a function of charge carrier density *n* and electric displacement field *D* for out-of-plane magnetic fields **(a)** $B_\perp$ = 1 T, **(b)** $B_\perp$ = 2 T and **(c)** $B_\perp$ = 3 T. Arrows mark the minima in *G* that correspond to phase transitions occurring at ν = ± 3 and a critical displacement field *D**. **(d)** Phase transitions as a function of $B_\perp$ and *D**. The dashed lines are fits to the data. The critical magnetic field at which the two fits cross is marked with B*.

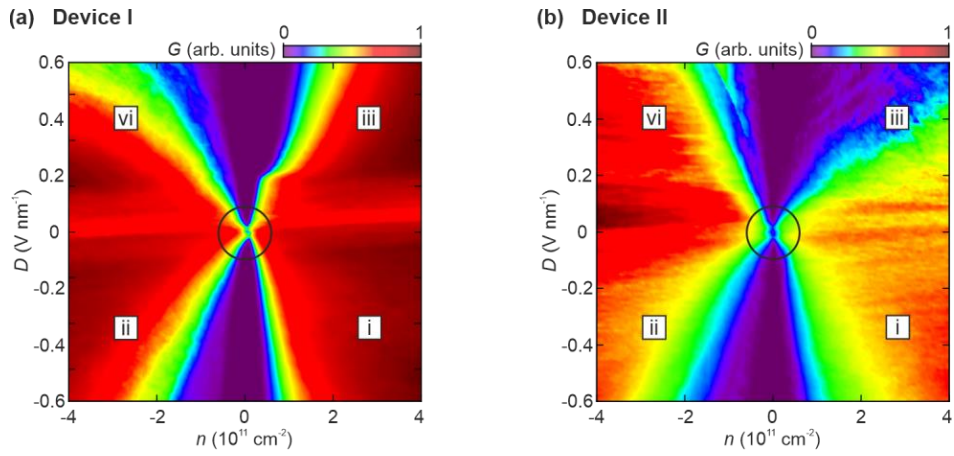

**Extended Data Fig. 3. Full phase diagrams of exchange and spin-orbit driven phases in layer-selective spin-orbit proximate bilayer graphene (a-b)** Conductance G in arbitrary units (arb. units) of charge carrier density $n$ and electric displacement field $D$ measured in Device I **(a)** and Device II **(b)** at zero magnetic field $B$. The labels i-iv mark the four quadrants in the $n$-$D$ plane (see Fig. 1). The insulating state at $n = D = 0$ is encircled.

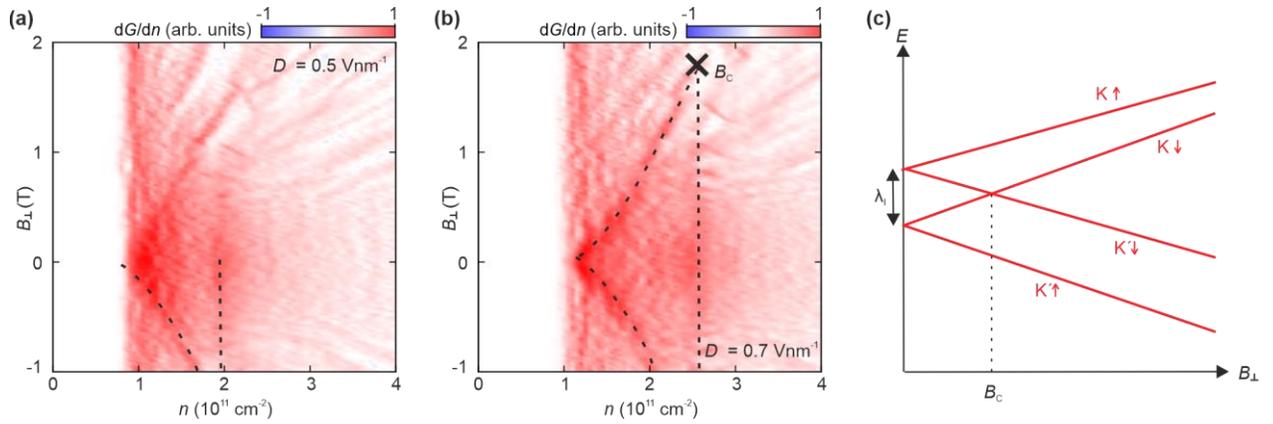

**Extended Data Fig. 4. Out-of-plane magnetic field dependence of electron-doped bilayer graphene with SOC. (a-b)** Derivative of the conductance d$G$/d$n$ as a function of charge carrier denisty $n$ and out-of-plane magnetic field $B_\perp$ measured in Device I for an electric displacement field $D$ = 0.5 V nm$^{-1}$ **(a)** and $D$ = 0.7 V nm$^{-1}$ **(b)**. Phase boundaries are highlighted by dashed lines. The phase boundaries shift to different values of $n$ for different applied $D$. **(c)** Energy dispersion as a function of $B_\perp$. The $B_\perp$-field induced crossing of the lowest spin-valley split band with the band of opposite spin polarization is marked as $B_C$.

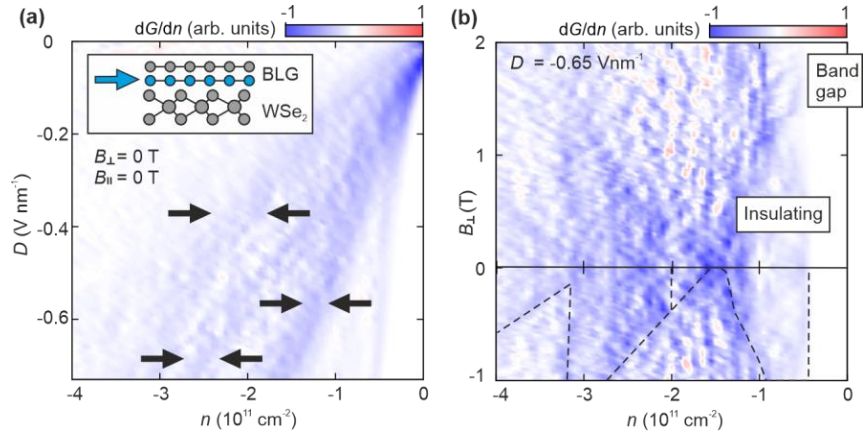

**Extended Data Fig. 5. Phase diagram of spin-orbit induced phases in hole-doped bilayer graphene. (a)** Derivative of the conductance d$G$/d$n$ as a function of charge carrier density $n$ and negative electric displacement fields $D$ measured in Device I. Phase boundaries are highlighted by arrows. **(b)** $G$ as a function of n and out-of-plane magnetic field $B_\perp$ at $D$ = -0.65 Vnm$^{-1}$. Phase boundaries are highlighted by dashed lines for negative $B_\perp$.

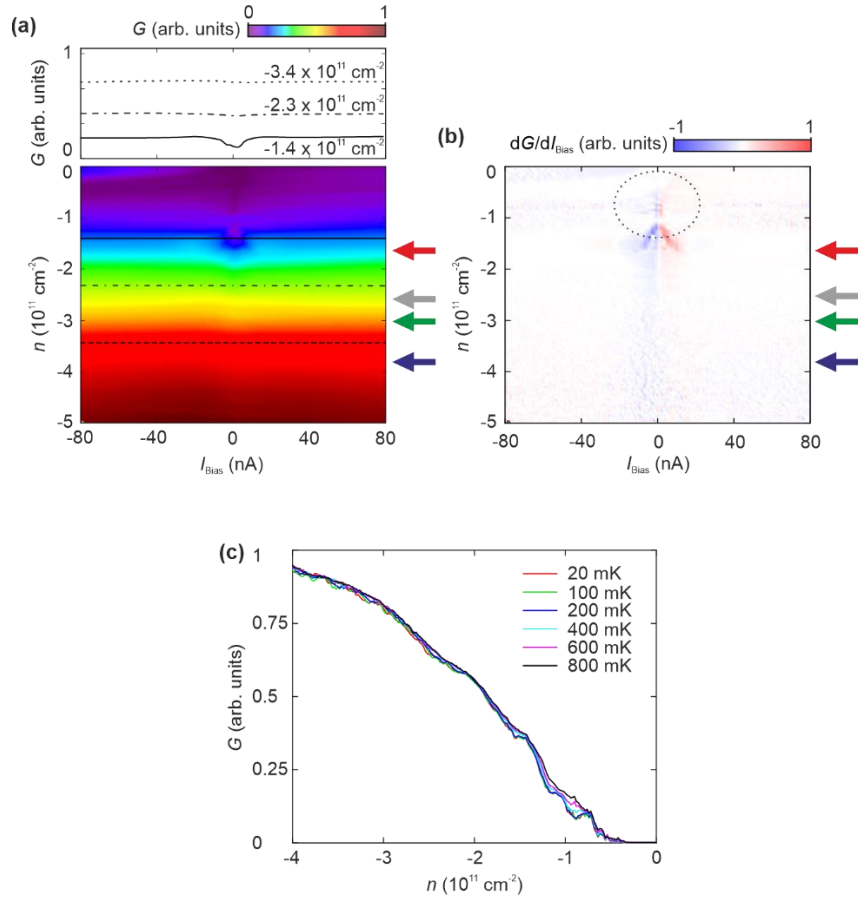

**Extended Data Fig. 6. Bias current and temperature dependences of the low-conductive regime in hole-doped bilayer graphene.** Charge transport commences here at the WSe$_2$-proximitized side. **(a,b)** Conductance $G$ **(a)** and derivative of the conductance (d$G$/d$I_{Bias}$) **(b)** as a function of the applied bias current $I_{Bias}$ and the charge carrier density $n$ for an electric displacement fields $D$ = -0.65 V nm$^{-1}$ measured in Device II. The regime of low conductance is encircled in **(b)**. Phase boundaries are marked by arrows. **(c)** Conductance $G$ as a function of n for different temperatures. The conductance exhibits minimal sensitivity to changes in temperature.

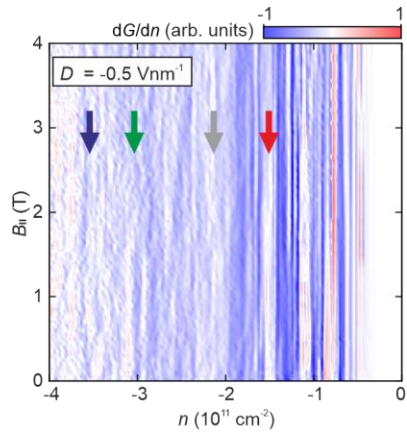

**Extended Data Fig. 7. In-plane magnetic field dependence of SOC induced phases in hole-doped bilayer graphene.** Derivative of the conductance d$G$/d$n$ as a function of $n$ and $B_{||}$ measured in Device I at $D$ = -0.5 V nm$^{-1}$ in the SOC-remote side of BLG. Phase boundaries are highlighted by arrows. Application of $B_{||}$ map up to 4 T does not change the phase boundaries since in this B-field regime $E_Z < \lambda_{SOC}$.

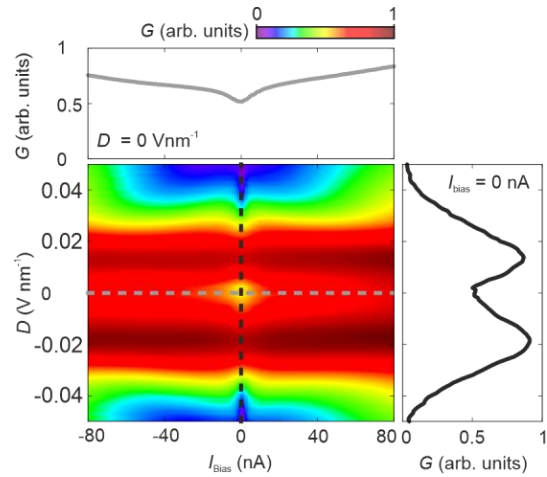

**Extended Data Fig. 8. Bias current dependence of the state at overall charge neutrality of the BLG/WSe$_2$ heterostucture.** Conductance $G$ as a function of the applied bias current $I_{Bias}$ and the electric displacement field $D$ at zero charge carrier density measured in Device I. The conductance at $D$ = 0 increases with increasing $I_{Bias}$, consistent with insulating behavior.

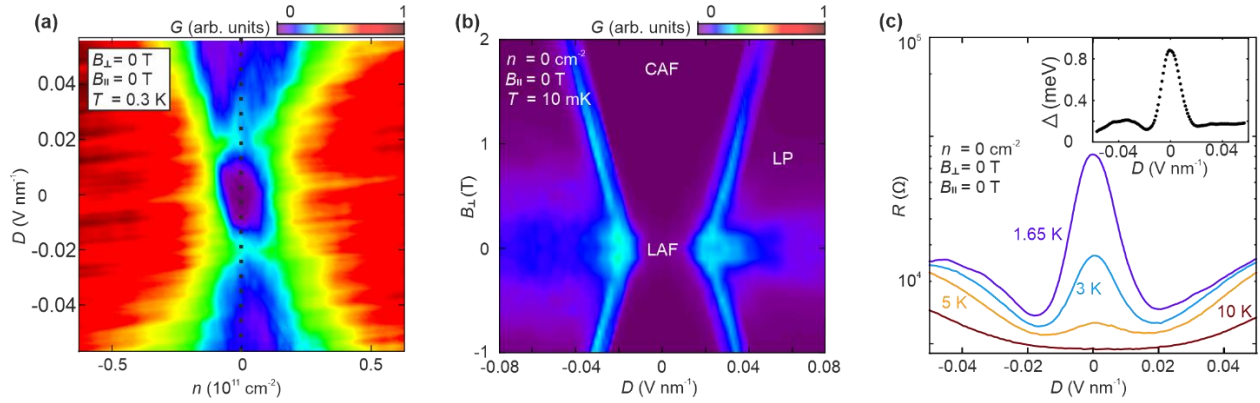

**Extended Data Fig. 9. Layer antiferromagnetic (LAF) state measured in freestanding bilayer graphene.**

**(a)** Two-terminal conductance $G$ in arbitrary units (arb. units) as a function of electric displacement field $D$ and charge carrier density $n$ at zero magnetic field $B$ and measured in Device III. **(b)** $G$ as a function of $D$ and an out-of-plane magnetic field $B_\perp$ at $n$ = 0. The layer antiferromagnetic (LAF), canted antiferromagnetic (CAF) and layer polarized (LP) states are labeled. **(c)** Two-terminal resistance $R$ as a function of $D$ for different temperatures. The inset shows the size of the activation gap $\Delta$ which was determined via Arrhenius fits, as a function of $D$. The data is comparable to the data measured in WSe$_2$ approximated bilayer graphene shown in Fig. 4.

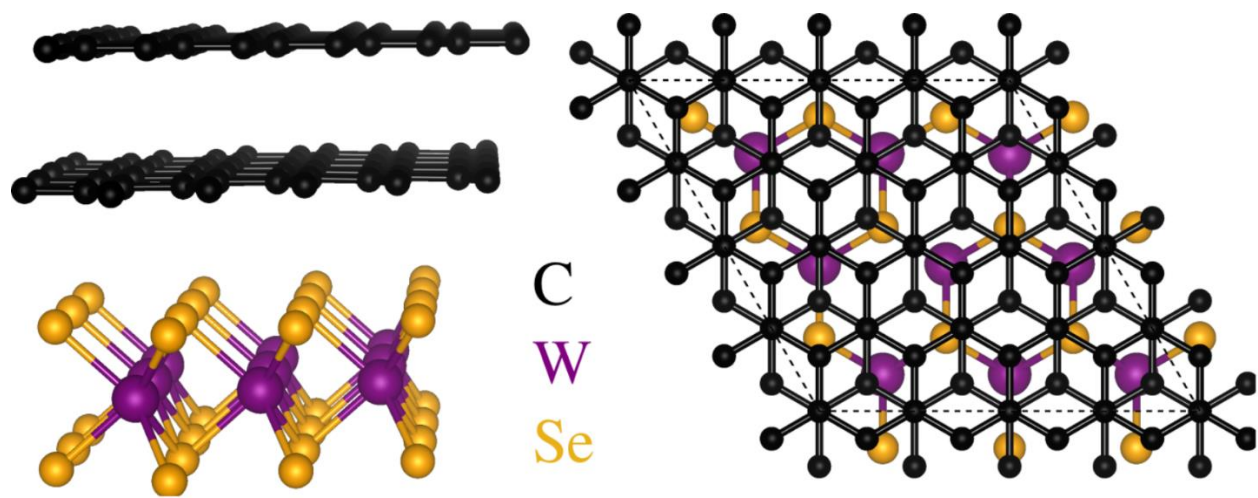

**Extended Data Fig. 10. Top and side view of the BLG/WSe₂ heterostructure.** The heterostructure unit cell for the DFT calculations has a lattice constant of 9.84 Å and consists of 91 atoms.

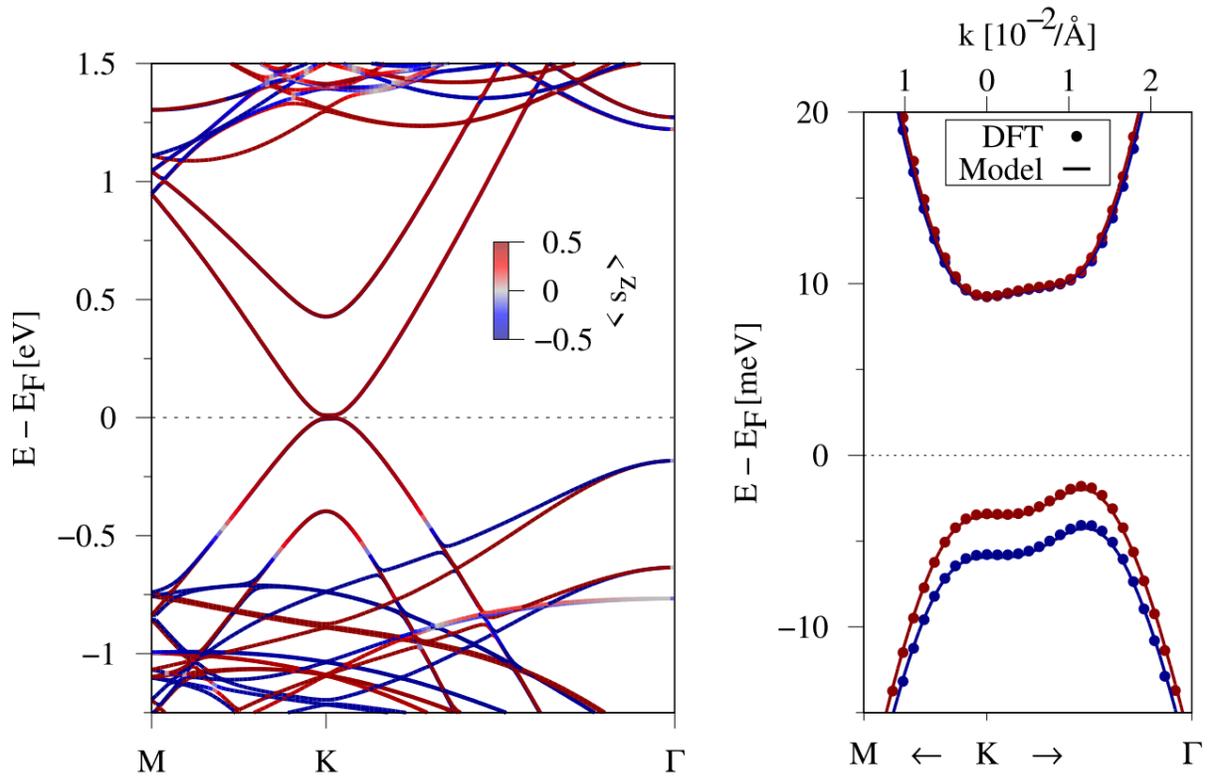

**Extended Data Fig. 11. Band structure of the BLG/WSe₂ heterostructure.** Left: DFT-calculated band structure of the BLG/WSe₂ heterostructure. The color code corresponds to the $s_z$ spin expectation value. Right: A zoom to the low energy bands near the Fermi level. The symbols correspond to the DFT data, while the solid lines represent the model Hamiltonian fit.

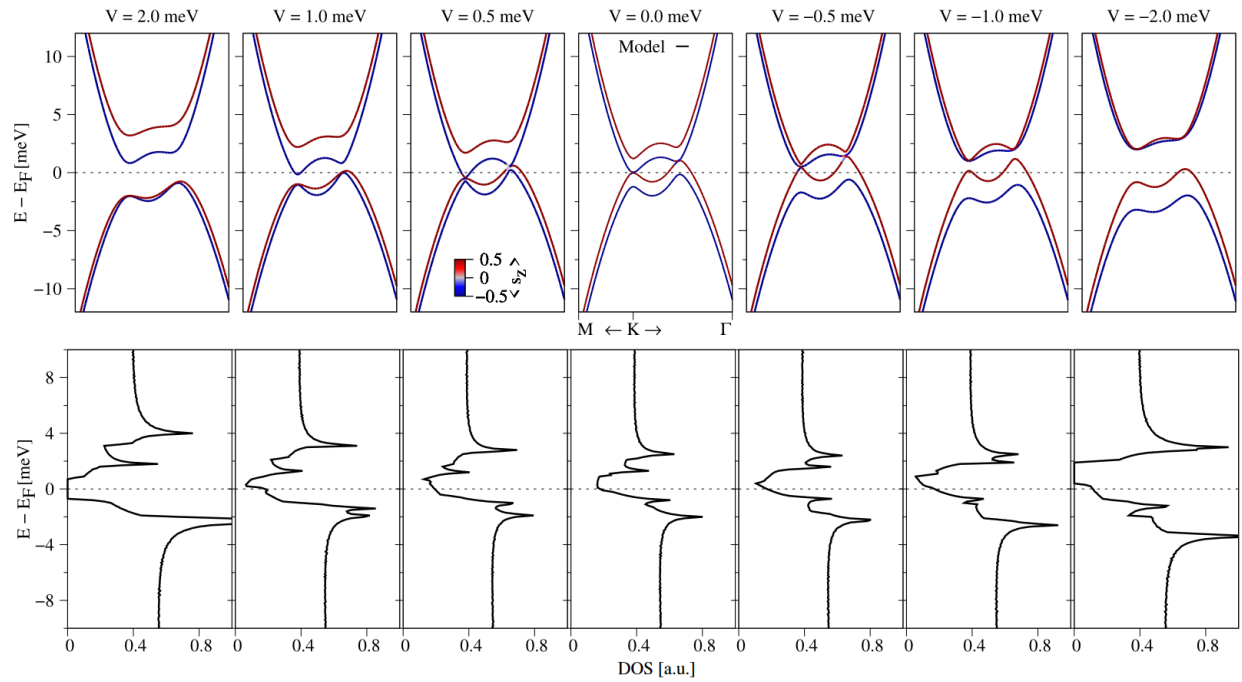

**Extended Data Fig. 12. Energy bands and density of states of the BLG/WSe₂ heterostructure.** The model-calculated low energy bands and the corresponding density of states of the BLG/WSe₂ heterostructure at various different interlayer electric potentials *V*.

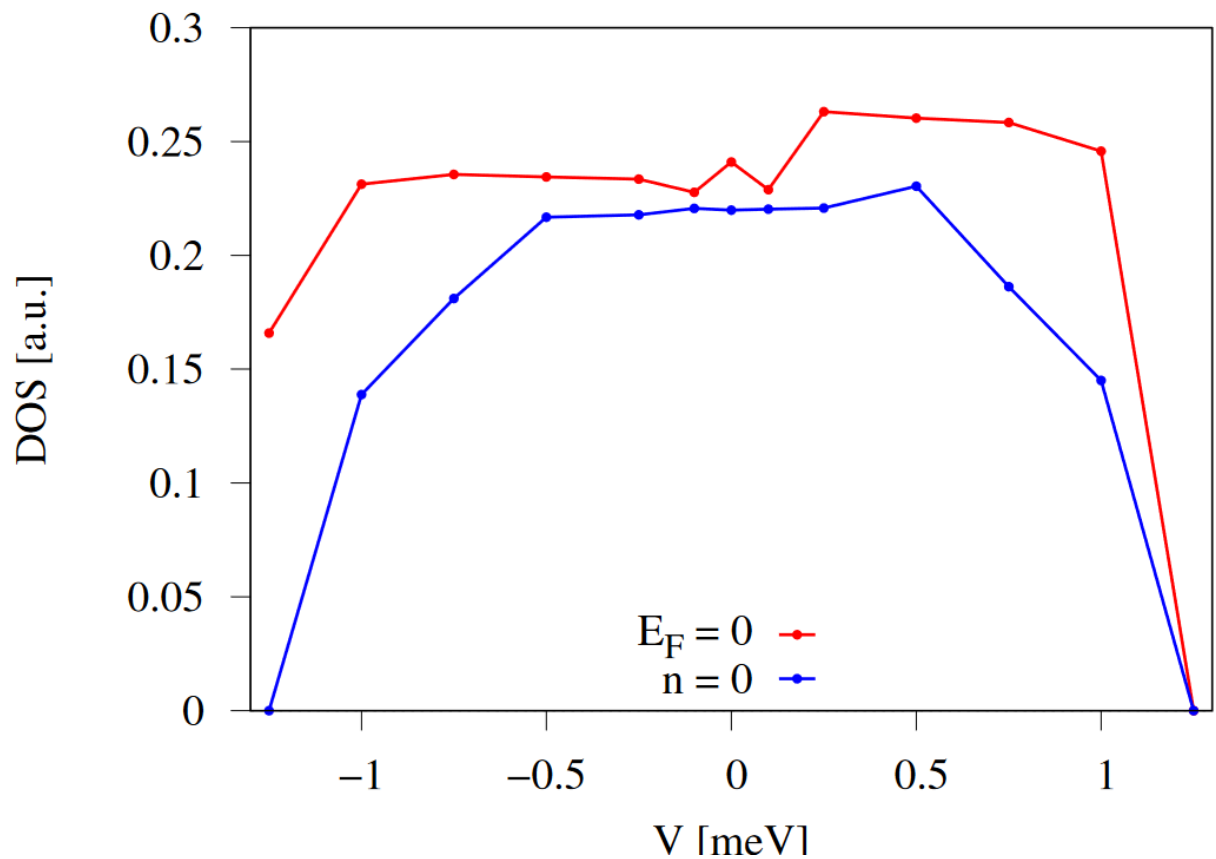

**Extended Data Fig. 13. Density of states of the BLG/WSe$_2$ heterostructure.** The model-calculated density of states (DOS) as a function of the interlayer potential *V*. $E_F = 0$ corresponds to the DOS at the Fermi level, while *n* = 0 corresponds to the DOS at charge neutrality.

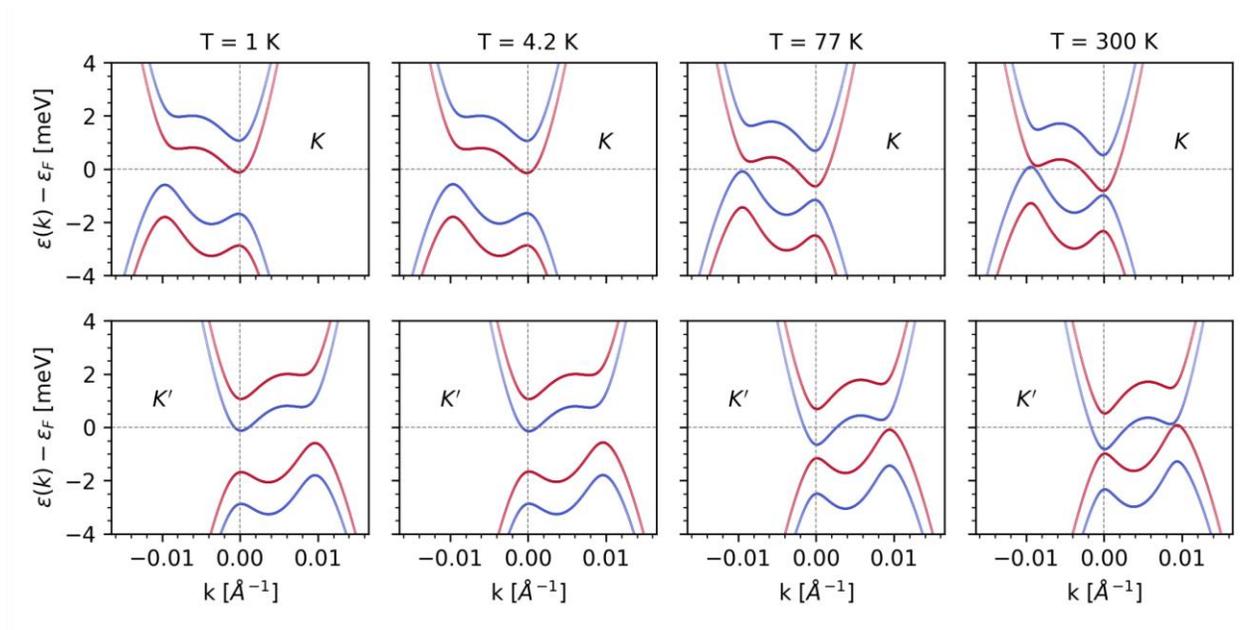

**Extended Data Fig. 14. Hartree-Fock calculations.** Temperature dependent Hartree Fock band structure of BLG/WSe$_2$ heterostructure at zero level of doping. Color code denote spin polarization: red, blue, and grey lines mark, correspondingly, spin-up, spin-down, and spin-unpolarized states.